# Acceleration of three-dimensional Tokamak magnetohydrodynamical code with graphics processing unit and OpenACC heterogeneous parallel programming

H. W. Zhang[1], J. Zhu[1], Z. W. Ma[1]*, G. Y. Kan[2]*, X. Wang[1], W. Zhang[1]

[1] Institute for Fusion Theory and Simulation, Zhejiang University, Hangzhou 310027, China

[2] State Key Laboratory of Simulation and Regulation of Water Cycle in River Basin, Research Center on Flood & Drought Disaster Reduction of the Ministry of Water Resources, China Institute of Water Resources and Hydropower Research, Beijing 100038, P. R. China

**Abstract:** In this paper, the OpenACC heterogeneous parallel programming model is successfully applied to modification and acceleration of the three-dimensional Tokamak magnetohydrodynamical code (CLTx). Through combination of OpenACC and MPI technologies, CLTx is further parallelized by using multiple-GPUs. Significant speedup ratios are achieved on NVIDIA TITAN Xp and TITAN V GPUs, respectively, with very few modifications of CLTx. Furthermore, the validity of the double precision calculations on the above-mentioned two graphics cards has also been strictly verified with m/n=2/1 resistive tearing mode instability in Tokamak.

* Author to whom correspondence should be addressed:

zwma@zju.edu.cn (Z. W. Ma); kanguangyuan@126.com (G. Y. Kan)

## 1. Introduction

Magnetic confinement fusion is a method of using a magnetic field to confine a high-temperature fusion fuel, deuterium-tritium plasma, to generate thermonuclear fusion energy. There are different kinds of magnetically confined fusion devices in operation or under construction around the world, mainly Tokamaks, such as the DIII-D Tokamak (Evans et al. 2005) in the U. S., the Joint European Torus (JET) Tokamak (Liang et al. 2007) in Europe, the EAST Tokamak (Wan 2009) in China, the Wendelstein 7-X (W7-X) Stellarator (Renner et al. 2000) in Germany, the International Thermonuclear Experimental Reactor (ITER) under construction (Bécoulet et al. 2008), and the China Fusion Engineering Test Reactor (CFETR) under design (Song et al. 2014). Because phenomena observed in these devices are too complex to be studied analytically, computational simulation becomes a powerful tool to investigate their inside physical mechanisms.

Considering the restrict of computational capabilities of the single processor core, parallel programming plays a major role in the basic research and application fields of program acceleration. Multiple parallel acceleration methods, such as MPI (Walker and Dongarra 1996) and OpenMP (Dagum and Menon 1998), have been developed in past decades. In recent years, benefitting from the rapid performance improvement of the graphics processing unit (GPU), new parallel programming methods and tools, such as CUDA (Cook 2012), OpenCL (Kaeli et al. 2015), and OpenACC (Farber 2016), have been developed vigorously.

OpenACC as a user-driven directive-based performance-portable parallel programming model, is developed to simplify the parallel programming for scientists and engineers. Compared with the CUDA and OpenCL which require great efforts on code redevelopment, OpenACC has many advantages, such as satisfactory acceleration with very few modifications on an original source code and good compatibility with other devices, for example central processing unit (CPU). It has been successfully applied in some scientific and engineering codes, such as the flow code NeK5000 (Markidis et al. 2015), the computational electromagnetics code Nekton (Otten et al. 2016), the C++ flow solver ZFS (Kraus et al. 2014), the Rational Hybrid Monte Carlo (RHMC) QCD code (Gupta and Majumdar 2018), the Gyrokinetic Toroidal Code (GTC) (Wang et al. 2016), the space plasma Particle-in-cell (PIC) code iPIC3D (Peng et al. 2015), the three dimensional pseudo-spectral compressible magnetohydrodynamic GPU code G-MHD3D (Mukherjee et al. 2018), the solar MHD code MAS (Caplan et al. 2018), and etc. The application perspective of the OpenACC technology in other scientific and engineering areas is very good and attractive.

CLT code (CLT is the abbreviation of Ci-Liu-Ti, which means magnetohydrodynamics in Chinese) is recently developed for the simulation study of magnetohydrodynamics (MHD) instabilities in toroidal devices, such as Tokamak (Wang and Ma 2015; Zhang, Ma, and Wang 2017). And it has been upgraded to CLTx in which the simulation domain extends to a scrape-off layer (SOL). The outstanding issue is the poor parallel computational efficiency in accelerating the MHD code with MPI and OpenMP. With implementation of OpenACC, satisfactory acceleration of CLTx with NVIDIA TITAN Xp and TITAN V GPUs is achieved. Compared with the speed of the MPI-parallelized CLTx executed on 64-core CPUs (Intel® Xeon® Gold 6148F), we get about four times of acceleration with single TITAN V and double TITAN Xp GPUs. In addition, the simulation result on GPU is also verified in accuracy.

The paper is organized as follows: In Section 2, the MHD model and the main modules of CLTx will be presented. In Section 3, the OpenACC implementation combined with MPI is given. And the acceleration performance of CLTx with OpenACC is analyzed in Section 4. The benchmark of simulation result from CLTx with implementation of OpenACC is presented in Section 5. Finally, conclusions and discussion are given in Section 6.

2. **Overview of CLTx**

CLTx is an initial value full MHD code with toroidal geometries to study MHD instabilities in magnetically confined fusion devices, such as Tokamak. The code is written in FORTRAN 90 with double precision format and has grown to more than 20,000 code lines with hundreds of subroutines and functions. CLTx has been successfully applied in studying the influence of toroidal rotation (Wang and Ma 2015), external driven current (Wang, Ma, and Zhang 2016; Zhang, Wang, and Ma 2017), and Hall effect (Zhang, Ma, and Wang 2017) on resistive tearing modes in Tokamaks. Its hybrid kinetic-magnetohydrodynamic version code, CLT-K, has also been developed to investigate dynamics of toroidal Alfvénic eigenmodes in Tokamak (Zhu, Ma, and Wang 2016; Zhu et al. 2018).

## 2.1 Basic MHD equations

The MHD simulation is frequently used in magnetic confined fusion. In Tokamak, the time scale of an MHD instability is much longer than that of microscopic particle dynamics, and its spatial scale is much larger than the particle gyroradius. In this circumstance, the particle-in-cell method is failure due to its unbearable computation requirements, and the fluid approximation becomes quite effective and workable.

Apart from the continuity, momentum and energy equations widely used in computational fluid dynamics (CFD), MHD equations contain the Lorentz force in the momentum equation, Faraday's law, generalized Ohm's law, and Ampere's law neglecting displacement current. Because of the similarities between CFD and MHD, CFD solvers can be generalized to study MHD problems (Khodak 2015), and vice versa.

The full set of resistive MHD equations including dissipations can be written as follows:

$$\partial_t \rho = -\nabla \cdot (\rho \mathbf{v}) + \nabla \cdot [D \nabla (\rho - \rho_0)], \quad (1)$$

$$\partial_t p = -\mathbf{v} \cdot \nabla p - \Gamma p \nabla \cdot \mathbf{v} + \nabla \cdot [\kappa (p - p_0)], \quad (2)$$

$$\partial_t \mathbf{v} = -\mathbf{v} \cdot \nabla \mathbf{v} + (\mathbf{J} \times \mathbf{B} - \nabla p) / \rho + \nabla \cdot [\nu (v - v_0)], \quad (3)$$

$$\partial_t \mathbf{B} = -\nabla \times \mathbf{E}, \quad (4)$$

with

$$\mathbf{E} = -\mathbf{v} \times \mathbf{B} + \eta (\mathbf{J} - \mathbf{J_0}) + \frac{d_i}{\rho} (\mathbf{J} \times \mathbf{B} - \nabla p), \quad (5)$$

$$\mathbf{J} = \nabla \times \mathbf{B}, \quad (6)$$

where $\rho$, $p$, $\mathbf{v}$, $\mathbf{B}$, $\mathbf{E}$, and $\mathbf{J}$ are the plasma density, thermal pressure, plasma velocity, magnetic field, electric field, and current density, respectively. $\Gamma (= 5/3)$ is the ratio of specific heat of plasma.

All variables are normalized as follows:

$\mathbf{B}/B_m \to \mathbf{B}$, $\mathbf{x}/a \to \mathbf{x}$, $\rho/\rho_m \to \rho$, $\mathbf{v}/v_A$, $t/\tau_a \to t$, $p/(B_m^2/\mu_0) \to p$,

$\mathbf{J}/(B_m/\mu_0 a) \to \mathbf{J}$, $\mathbf{E}/(v_A B_m) \to \mathbf{E}$, and $\eta/(\mu_0 a^2/\tau_a) \to \eta$,

where $\tau_a = a/v_A$ is the Alfvénic time; $v_A = B_m/(\mu_0 \rho_m)^{1/2}$ is the Alfvénic speed; $B_m$ and $\rho_m$ are the magnetic field and plasma density at the magnetic axis, respectively; $a$ is the minor radius in the poloidal cross-section.

Constrained by the equilibrium conditions, the following equations should be satisfied:

$$\nabla \cdot (\rho_0 \mathbf{v}_0) = 0, \tag{7}$$

$$\mathbf{v}_0 \cdot \nabla p_0 + \Gamma p_0 \nabla \cdot \mathbf{v}_0 = 0, \tag{8}$$

$$\rho_0 \mathbf{v}_0 \cdot \nabla \mathbf{v}_0 = \mathbf{J}_0 \times \mathbf{B}_0 - \nabla p_0, \tag{9}$$

$$\nabla \times \mathbf{E}_0 = 0. \tag{10}$$

Substituting these equilibrium equations into Equations (1)-(4), the MHD equations can be rewritten as

$$\partial_t \rho = -\nabla \cdot (\rho \mathbf{v}_1 + \rho_1 \mathbf{v}_0) + \nabla \cdot [D \nabla (\rho - \rho_0)], \tag{11}$$

$$\partial_t p = -\mathbf{v}_1 \cdot \nabla p - \mathbf{v}_0 \cdot \nabla p_1 - \Gamma (p \nabla \cdot \mathbf{v}_1 + p_1 \nabla \cdot \mathbf{v}_0) + \nabla \cdot [\kappa (p - p_0)], \tag{12}$$

$$\partial_t \mathbf{v} = -(\mathbf{v} \cdot \nabla \mathbf{v}_1 + \mathbf{v}_1 \cdot \nabla \mathbf{v}_0 + \rho_1 \mathbf{v}_0 \cdot \nabla \mathbf{v}_0 / \rho)$$
$$+ (\mathbf{J}_1 \times \mathbf{B} + \mathbf{J}_0 \times \mathbf{B}_1 - \nabla p_1)/\rho + \nabla \cdot [\nu (v - v_0)], \tag{13}$$

$$\partial_t \mathbf{B} = -\nabla \times \mathbf{E}_1, \tag{14}$$

where the variables with subscript 0 represent equilibrium components and 1 for perturbation components, e.g., $\mathbf{v}_1 = \mathbf{v} - \mathbf{v}_0$. Thus, numerical errors from equilibrium can be minimized.

## 2.2 Coordinate systems and numerical schemes

In CLTx, the cylindrical coordinate system $(R, \phi, Z)$ as shown in Figure 1 is adopted. The three-dimensional simulation boundary determined by the last closed magnetic surface and the m/n=3/1 magnetic island for the EAST Tokamak from CLTx are also plotted in Figure 1. In the magnetic confinement fusion device such as Tokamak, $R$, $\phi$ and $Z$ indicate the major radius, toroidal, and up-down directions, respectively. One advantage of this coordinate system is that we can avoid the singularity near the $r=0$ point that occurs in the toroidal coordinate $(\psi,\theta,\zeta)$. However, the boundary handling would be more difficult in the cylindrical coordinate. In the previous version of CLTx, the fixed boundary condition at the last flux surface of plasma is assumed. Recently, the cut-cell method (Duan, Wang, and Zhong 2010) has been applied in CLTx successfully, the details of the new boundary handling method will be introduced in another paper.

The uniform and rectangular grid is used in the $R-Z$ directions. For the numerical scheme, the 4$^{th}$ order finite difference method is employed in the $R$ and $Z$ directions, while in the $\phi$ direction, either 4$^{th}$ order finite difference or pseudo-spectrum method can be used. As for the time-advance, the 4$^{th}$ order Runge-Kutta scheme is applied.

For the parallelization on CPU platform, the simulation domain can be divided into multiple blocks in each direction, and four-layer grids are used for message passing between every two neighboring blocks according to the 4$^{th}$ order finite difference method in space. Thus, the increase of MPI CPU cores with a fixed mesh size will lead to rapid deterioration of the acceleration efficiency.

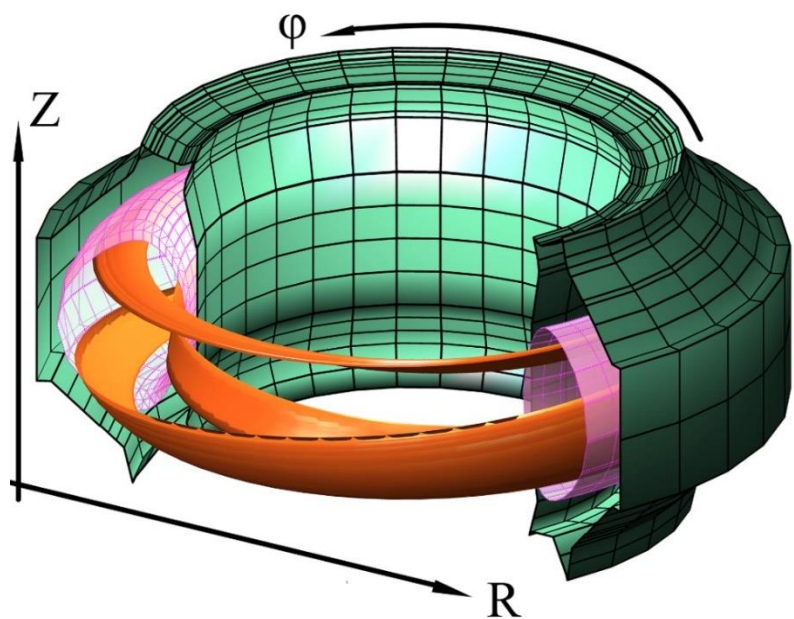

**Figure 1.** A three-dimensional perspective view of magnetic surface geometry for the EAST Tokamak. The three brown ribbons indicate m/n=3/1 magnetic island from CLTx. The pink surface represents the last closed magnetic surface while the green surface is the simulation boundary. The adopted cylindrical coordinate system $(R,\varphi,Z)$ is also given.

## 2.3 Main modules and basic execution flow chart of CLTx

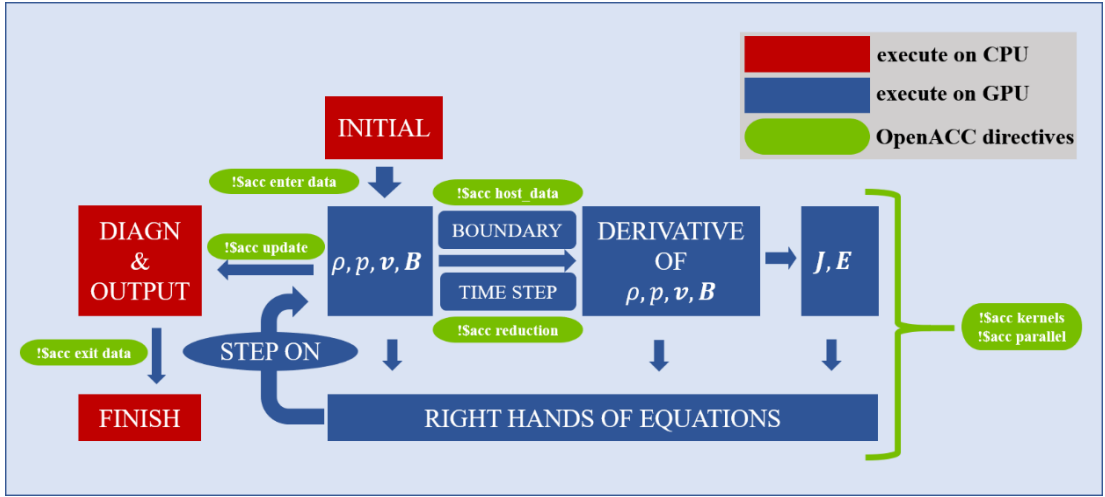

**Figure 2.** Main modules and OpenACC directives of CLTx. The modules in the blue blocks are executed on GPU, and those in the red blocks are on CPU. The green blocks indicate the OpenACC directives applied in the corresponding positions.

For each iterative step of CLTx, the eight components of $\rho, p, v, B$ need to be updated. The main processes of CLTx are shown in Figure 2. The modules with red blocks are executed on CPU, while the blue ones on GPU, and the green blocks mark out the main OpenACC directives in the whole procedure. The **INITIAL** module reads the equilibrium data that is calculated by the equilibrium solving code, such as NOVA (Cheng and Chance 1987) and EFIT (Lao et al. 1985). Then, the **DERIVATIVES OF** $\rho, p, v, B$ module using the 4$^{th}$ order finite difference method is called to obtain the derivatives. With these derivatives, the values of the current $J$ and electric field $E$ are obtained. The derivatives, current $J$ and electric field $E$, are substituted into the **RIGHT HANDS OF EQUATIONS**, and finally $\rho, p, v, B$ are updated in **STEP ON** module with the 4$^{th}$ order Runge-Kutta scheme. Note that during the iteration the **BDOUNDARY** module is applied in each step to solve the boundary conditions and deals with data exchange due to MPI parallelization, while the **TIME STEP** module calculates the time increment based on the latest $\rho, p, v, B$ to satisfy the Courant-Friedrichs-Lewy (CFL) condition in whole simulation domain, and some other modules such as **DIAGN** (recording data for time sequence analysis) and **OUTPUT** (recording data for global analysis) are called every few steps.

The execution time of CLTx is mostly spent on the **RIGHT HANDS OF EQUATIONS**, **DERIVATIVES OF** $\rho, p, v, B$, and $J, E$ modules. Therefore, a majority of OpenACC directives are added in these modules.

## 3. OpenACC and MPI implementations in CLTx

### 3.1 OpenACC directives added to the serial version of CLTx

The main idea of GPU-acceleration using OpenACC is to offload the calculation process from CPU to GPU. The data transfer speed between the memories of CPU and GPU is much slower than that of GPU self-memory access. Thus, the first important operation is to copy all variables required in iterations from CPU memory to GPU memory. Therefore, after reading the equilibrium data in the **INITIAL** module, all necessary variables are copied into the GPU memory by using *'$acc enter data copyin (variable name list)'* directive and clause as shown in Figure 2. Accordingly, after finishing all iterations, the *'$acc exit data delete (variable name list)'* directive and clause are also used before the program finishes to free the GPU memory. There are also other methods playing the same role, such as combination of *'!$acc declare create (variable name list)'* and *'!$acc update device (variable name list)'*. The main advantage of *'enter data'* and *'declare create'* construct is that the data lifetime in the GPU memory can be extended among different functions and subroutines. With this feature, the calculation on the GPU can be continuously proceeded without stopping for data transfer between CPU and GPU.

A demonstrative example for the *'enter data'* constructs in CLTx is given in Figure 3. For simplicity, only three representative variables ($xx, yy, zz$) are listed in the OpenACC *'copyin'* clause. During the main iteration, CLTx calls the **DIAGN** and **OUTPUT** modules every few steps to record data into hard disk by CPU. Because the **DIAGN** module must use some external math libraries, such as MKL (Wang et al. 2014), this module is still executed on CPU due to its difficulty for parallelization with OpenACC. Therefore, just before calling the **DIAGN** and **OUTPUT** modules, the directive *'!$acc update host (variable name list)'* is used to copy the latest data from GPU back to CPU. The time wasted on updating operation is negligible because the number of calls of these two modules is rather limited.

```fortran
PROGRAM CLTx
! INITIALIZE
CALL INITIA
!$acc enter data copyin(xx, yy, zz)
! MAIN ITERATION PART
DO NSTEP = 1, NSTEP
    CALL STEPON
    IF(MOD(NSTEP,NDIAGN) == 0) THEN
        !$acc update host(xx, yy, zz)
        CALL DIAGN
    ENDIF
    IF(MOD(NSTEP,NOUTPUT) == 0) THEN
        !$acc update host(xx, yy, zz)
        CALL OUTPUT
    ENDIF
ENDDO
!$acc exit data delete(xx, yy, zz)
END PROGRAM CLTx
```

**Figure 3.** An example for *'enter data'* and *'update'* constructs in CLTx.

Since CLTx is a three-dimensional MHD code, the majority of loops contains three levels. To parallelize the loops, two constructs of *'kernels'* and *'parallel'* are frequently used. A representative example for such parallelization in the **TIME STEP** module is given in Figure 4. The *'present'* clause following *'parallel'* construct directive indicates that the variables inside the parentheses already exist on GPU, which can avoid the needless implicit variable copy or create operations on GPU. The *'loop'* clause after *'parallel'* directive should be placed just before the loop body, which instructs the compiler to perform parallelization for the closely followed loop. As for the **TIME STEP** module, the reduction for the minimum time step is used by adding *'reduction(min:dt1)'*, because the time step used in time advance must satisfy the CFL condition in each grid. Other reduction operations are also supported in OpenACC with similar syntaxes (Farber 2016). The *'independent'* clause instructs the compiler that the calculation on each grid has no dependency and therefore can be parallelized directly. The final clause in Figure 4 is *'collapse(3)'*. Collapsing loops means that, for example, three loops of trip counts NX, NZ, and NY will automatically combine into a single loop with a trip total count of NX*NZ*NY, which greatly increases the parallelization efficiency. Note that the *'parallel'* constructs used here can also be replaced by *'kernels'* construct, the major difference of these two constructs is that *'kernels'* gives the compiler more freedom to find and map parallelism according to the requirements of the target accelerator, while the parallel construct is more explicit, and requires more analysis by the programmer.

Another important OpenACC directive used in CLTx is the *'!$acc routine'* directive for the procedure call. Figure 5 gives the solution when a subroutine or function is called inside an accelerated region. The subroutine *'interp1d2l'* interpolates data between different mesh grids, and is commonly used in the accelerated region of **BOUNDARY** module. The *'!$acc routine seq'* directive instructs the compiler to generate a device version for this subroutine on GPU so that it can be called in the accelerated region. An interface for this child subroutine is required inside the parent as given in Figure 5. Then the subroutine can be called inside the OpenACC accelerated region directly.

```fortran
SUBROUTINE SETDT
USE DECLARE
INCLUDE 'MPIF.H'
DT1=100.d0
!$acc parallel present(x, xx, yy, zz, gdtp_ep)
!$acc loop reduction(min:dt1) independent collapse(3)
DO JY = IY_FIRST + 2, IY_LAST - 2
DO JZ = IZ_FIRST + 2, IZ_LAST - 2
DO JX = IX_FIRST + 2, IX_LAST - 2
    IF(GDTP_EP(JX,JZ,1).NE.4) THEN
        VX=X(JX,JZ,JY,3)
        VY=X(JX,JZ,JY,4)
        VZ=X(JX,JZ,JY,5)
        VA2=(X(JX,JZ,JY,6)**2+X(JX,JZ,JY,7)**2  &
            +X(JX,JZ,JY,8)**2)/X(JX,JZ,JY,1)
        CS2=GAMMA*X(JX,JZ,JY,2)/X(JX,JZ,JY,1)

        VPX=DABS(VX)+SQRT(DABS(CS2+VA2))
        VPY=DABS(VY)+SQRT(DABS(CS2+VA2))
        VPZ=DABS(VZ)+SQRT(DABS(CS2+VA2))

        DTX=DABS(XX(JX)-XX(JX-1))/(VPX/CFL)
        DTZ=DABS(ZZ(JZ)-ZZ(JZ-1))/(VPZ/CFL)
        DTY=DABS(XX(JX)*(YY(JY)-YY(JY-1)))/(VPY/CFL)

        DT2=DMIN1(DTX,DTZ)
        DT3=DMIN1(DTY,DT2)
        DT1=DMIN1(DT1,DT3)
    ENDIF
ENDDO
ENDDO
ENDDO
!$acc end parallel
RETURN
END SUBROUTINE SETDT
```

**Figure 4.** An example for the parallelization in the **TIME STEP** module using *'parallel'* constructs.

```fortran
SUBROUTINE INTERP1D2L(X1, X2, X3, Y1, Y2, Y3, Y, ANS)
!$acc routine seq
REAL*8 X1,X2,X3,Y1,Y2,Y3,Y,ANS
REAL*8 D1,D2,D3
D1 = (Y1-Y2)*(Y1-Y3)
D2 = (Y2-Y3)*(Y2-Y1)
D3 = (Y3-Y1)*(Y3-Y2)
ANS = X1*(Y-Y2)*(Y-Y3)/D1 &
    + X2*(Y-Y3)*(Y-Y1)/D2 &
    + X3*(Y-Y1)*(Y-Y2)/D3

RETURN
END SUBROUTINE INTERP1D2L
!!!!!!!!!!!!!!!!!!!!!!!!!!!!!!!!!!!!!!!!!!!!!!!!!!!!!!
SUBROUTINE BOUNDARY
USE DECLARE
IMPLICIT NONE
INCLUDE 'MPIF.H'

! INTERFACE FOR ACCELERATED SUBROUTINE
INTERFACE
    SUBROUTINE INTERP1D2L(X1, X2, X3, Y1, Y2, Y3, Y, ANS)
    !$acc routine seq
    REAL*8 X1,X2,X3,Y1,Y2,Y3,Y,ANS
    REAL*8 D1,D2,D3,TMP_ADD
    END SUBROUTINE
END INTERFACE

! OpenACC ACCELERATED REGION
......
CALL INTERP1D2L(X1, X2, X3, Y1, Y2, Y3, Y, ANS)
......
! END OF OpenACC ACCELERATED REGION

END SUBROUTINE BOUNDARY
```

**Figure 5.** An example for the procedure call (subroutine interp1d2l) inside OpenACC accelerated region.

## 3.2 Combination of OpenACC and MPI technologies for parallelization on multiple GPUs

CLTx was developed with MPI at first for parallelization in clusters. For a case with (256,

256, 32) grids in the $(R, Z, \phi)$ directions, the GPU memory usage in a graphics card is about 2.25GB. Thus, the memory in a single graphics card such as TITAN Xp or TITAN V (with 12GB of GPU memory) is more than enough. However, for a high-resolution simulation, a size of (1024, 1024, 128) grids or even larger should be chosen. In such scenarios, more than 64 times memory (over 144GB) will be required on the GPU. Naturally, the multiple GPUs parallelization becomes indispensable. Based on the original MPI in CLTx, we also added the directives for directly communication between GPUs without transferring data back to CPU.

As shown in Figure 6, firstly, we add the *'!\$acc set device_num (rank of GPU)'* directive just after the *'mpi_comm_size'* and *'mpi_comm_rank'* calls. The value of *'rank of GPU'* can be defined as functions of the process rank in MPI parallelization so that each MPI rank can be assigned to a specified GPU for the load balance. With this *'rank of GPU'*, the communication between different GPUs becomes feasible. Then, a specific example with OpenACC directives in *'mpi_send'* and *'mpi_recv'* API is given. The *'!\$acc host_data use_device (variable name list) if_present'* instructs the compiler to use the address of the specified variables on GPU if the variables are present, then the *'mpi_send'* or *'mpi_recv'* API is executed by accessing the data from the GPU address of variable rather than the CPU. The adoption of the *'host_data use_device'* construct can avoid the unnecessary data transfer between CPU and GPU before MPI calls. For the purpose of comparison, an equivalent example is given in Figure 7, where the MPI still accesses the data from the CPU address for specified variables. However, in order to update the data inside the CPU and GPU, another pair of *'!\$acc update host (variable name list)'* and *'!\$acc update device (variable name list)'* are added before the *'mpi_send'* and after the *'mpi_recv'*, respectively. The functions of the methods used in Figure 6 and Figure 7 are the same, but the later one will spend additional time on data updating between CPU and GPU.

```fortran
PROGRAM CLTx
......
! INITIATE MPI
CALL MPI_INIT(IERROR)
CALL MPI_COMM_SIZE(MPI_COMM_WORLD,NSIZE,IERROR)
CALL MPI_COMM_RANK(MPI_COMM_WORLD,NRANK,IERROR)
!$acc set device_num(0)
......
!MPI SEND------------------------------------------------
!$acc host_data use_device(wfx1) if_present
CALL MPI_SEND(WFX1,MYZ8,MPI_DOUBLE_PRECISION, NRANK+1, 0, &
        MPI_COMM_WORLD, IERROR)
!$acc end host_data
......
!MPI RECEIVE---------------------------------------------
!$acc host_data use_device(wfx1) if_present
CALL MPI_RECV(WFX1,MYZ8,MPI_DOUBLE_PRECISION, NRANK-1, 0, &
        MPI_COMM_WORLD, IERROR)
!$acc end host_data
......
END PROGRAM CLTx
```

**Figure 6.** An example for MPI called on GPU by using *'host_data use_device'* construct.

```fortran
PROGRAM CLTx
......
! INITIATE MPI
CALL MPI_INIT(IERROR)
CALL MPI_COMM_SIZE(MPI_COMM_WORLD,NSIZE,IERROR)
CALL MPI_COMM_RANK(MPI_COMM_WORLD,NRANK,IERROR)
!$acc set device_num(0)
......
!MPI SEND------------------------------------------------
!$acc update host(wfx1)
CALL MPI_SEND(WFX1,MYZ8,MPI_DOUBLE_PRECISION, NRANK+1, 0, &
        MPI_COMM_WORLD, IERROR)
......
!MPI RECEIVE---------------------------------------------
CALL MPI_RECV(WFX1,MYZ8,MPI_DOUBLE_PRECISION, NRANK-1, 0, &
        MPI_COMM_WORLD, IERROR)
!$acc update device(wfx1)
......
END PROGRAM CLTx
```

**Figure 7.** An example for MPI called on CPU by using *'update'* construct.

## 4. Acceleration performance analysis

**Table 1.** Detailed information of two platforms used in the acceleration performance tests.

| platform | CPU configuration | memory configuration | GPU configuration | Fortran compiler version | MPI version | compiler options |
|---|---|---|---|---|---|---|
| **KYLIN-2 cluster** CentOS Linux release 7.4.1708 | 138 nodes, with 2 Intel® Xeon® Gold 6148F CPUs (i.e., 40 cores) in each node | 92GB in each node | N/A | Intel(R) Fortran Intel(R) 64 Compiler XE for applications running on Intel(R) 64, Version 15.0.3.187 Build 20150407 | mpiifort for the Intel(R) MPI Library 5.0 Update 3 for Linux* | -O3, -mcmodel=large |
| **x3860-x6 workstation** Red Hat Enterprise Linux Server release 6.8 | 1 node, with 16 Intel Xeon E7-8867 v4 CPUs (i.e., 144 cores in total) | 1024GB in total | 2 TITAN Xps & 2 TITAN Vs | pgfortran 18.10-0 64-bit target on x86-64 Linux -tp haswell PGI Compilers and Tools | mpif90, OpenMPI 2.1.2 | -acc, -mcmodel=medium, -Mlarge_arrays, -ta=tesla:cc70 (for TITAN V) & -ta=tesla:cc60 (for TITAN Xp) |

The performance tests are carried out on two platforms for CPU and GPU version of CLTx, respectively. The detailed information for each platform is given in Table 1, including the hardware configurations, compiler information, compiler options, and etc. The CPU version of CLTx is executed on KYLIN-2 cluster with Intel® Fortran compiler included in the Intel® Parallel Studio XE, while the GPU version is executed on x3860-x6 workstation with 2 TITAN Xps and 2 TITAN Vs installed, the compiler is pgfortran 18.10 included in the PGI Compilers and Tools. Note that the compiler options are different for TITAN Xp and TITAN V due to their different computational capabilities. And the OpenMPI on GPU platforms is supported by the PGI Compilers and Tools, which is used in the combination of OpenACC and MPI technologies for parallelization on multiple GPUs as discussed above.

### 4.1 Performance on single GPU card

For the case with grids as (256, 256, 16) in $R$, $Z$, and $\phi$, the comparison for execution time (with 20,000 steps) is carried out on different platforms. Some simulation results are presented in Figure 8. The MPI block division for the CPU version of CLTx is carried out uniformly in the $R$ and $Z$ directions, while no block division is applied in the GPU version because we only used one card for acceleration. Note that for the case executing on the CPU, the 'O3' compiler option is turned on so as to obtain the best performance optimization.

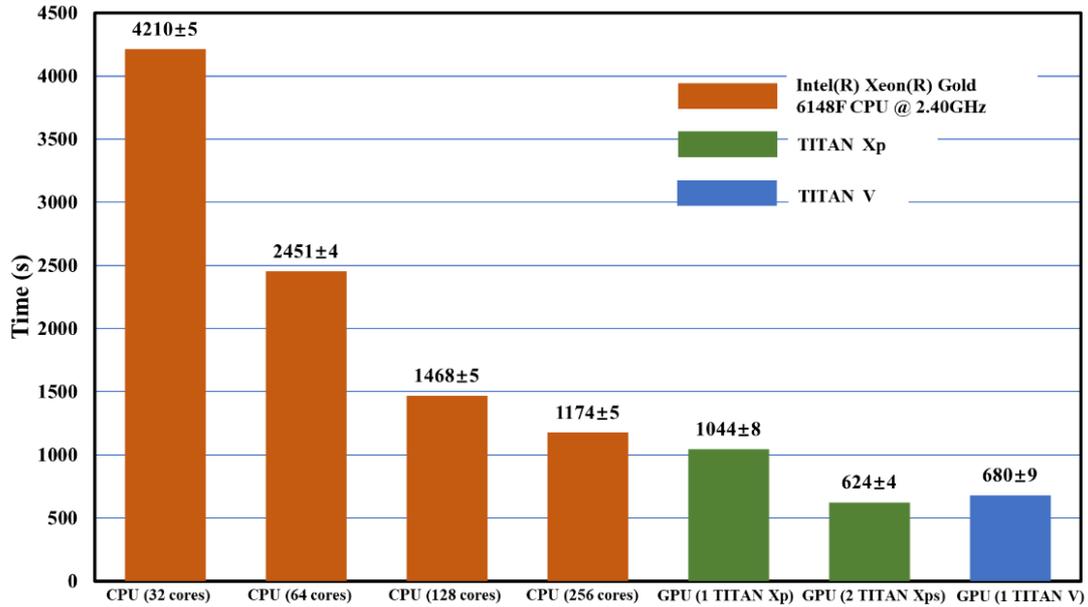

**Figure 8.** Comparison for execution time on different platforms (CPU colored by orange, TITAN Xp by green, and TITAN V by blue).

As shown in Figure 8, the performance of the code based on MPI executed on CPU is still acceptable when total 128 cores (4 nodes) are used, with the parallel acceleration efficiency at 72%. However, the efficiency drops quickly to 45% when the number of CPU cores is increased to 256 (7 nodes). The reason for the decline of the parallel acceleration efficiency with increase of CPU cores is mainly due to the growth of the overlap area between different MPI blocks with 4$^{th}$ order finite difference scheme in space. Therefore, the speedup for an MHD code with a fixed mesh is quite difficult with traditional MPI. Fortunately, the application of OpenACC based on GPU on our CLTx can achieve a quite satisfactory acceleration efficiency. As the results shown in Figure 8 with the green and blue bars for TITAN Xp and TITAN V, respectively, the elapsed time on single TITAN Xp almost equals to that of 256 Intel® Xeon® Gold 6148F CPU cores. What's more, the speed of TITAN V is even 54% faster than that of TITAN Xp. Compared with our frequently-used MPI set of 64 cores, the speedup of TITAN V is about 3.6 times. In addition to the speed performance, another advantage for the GPU acceleration is the much lower price for constructing GPU workstations or servers than building a cluster with CPU nodes. The cost of a single TITAN V is comparable with one Intel® Xeon® Gold 6148F CPU (20 cores in each). Thus, the cost performance of the former is about ten times better than the latter for our CLTx.

Another important feature used in OpenACC is the combination of three levels parallelization as *'gang'*, *'work'*, and *'vector'*. For the case shown in Figure 8, we found that the speed of CLTx on the specified mesh is most sensitive to the number of *'gang'*: enough gang parallelization (more than 512) leads to the best performance on GPU, while too few gangs (less than 100) results in slowing down of the code compared with the MPI case with 32 cores. The default set of the compiler without adding any artificially specified configurations also leads to the best performance of the acceleration as shown in Figure 8. The numbers of *'gang'*, *'worker'*, and *'vector'* will affect the usage of GPU memory cache and can influence the speed greatly. For an MHD code accelerated with OpenACC, adjusting these configurations is necessary to obtain the best acceleration performance, and the best set is also quite different for different codes, devices, and problem sizes.

### 4.2 Performance on double GPU cards combined with MPI

For the lack of cluster with multiple GPU nodes, the speed test for the OpenACC combined with MPI has only been performed on up to two separate GPU cards inside a workstation. Note that the computation and communication in CLTx is separated, that is, the synchronous communication method is adopted.

Firstly, a simple test for MPI-GPU combined acceleration is carried out with 4 MPI ranks on single TITAN Xp with the two implementation methods mentioned in Figure 6 and 7 (by setting 4 MPI threads in a same GPU rank). The first method (as shown in Figure 6, with direct communication between GPUs) leads to only about 15% slowing down compared with time showed by the green bar in Figure 8, while the other method (as shown in Figure 7, using *'update'* clause and communication between CPU MPI threads) leads to about 105% slowing down. Therefore, the support of direct MPI communication on GPU makes the multiple GPU acceleration for large size problems feasible, while the data exchange between GPU and CPU memories will intolerably slow down the execution speed.

Secondly, a corresponding test for MPI-GPU combined acceleration is performed on two TITAN Xp GPUs. The method of direct communication between GPUs as shown in Figure 6 is chosen. Two MPI processes in total are carried out with double TITAN Xp GPUs, that is, each GPU executes one MPI process, and the computational domain is equally divided into two blocks in the $R$ direction. As the result shown in Figure 8, the final execution time is about 624 seconds, which is even faster than that on single TITAN V. Compared with the performance of single TITAN Xp, the parallel efficiency is about 84%. The communication time is relatively small compared with the computation time, though the communication in CLTx is synchronous. Therefore, the time cost on data interaction through Peripheral Component Interface Express (PCIe) is negligible and the parallelization with multiple GPUs is workable, which is necessary and indispensable for simulating larger size problem in future research.

In order to further examine the parallel scalability of CLTx, the execution times for different mesh size $N_R$ in the $R$ direction in single and double TITAN Xps are studied and given in Figure 9 ($N_R$ equals to 32, 64, 128, 256 and 512, respectively, while $N_Z$ and $N_\phi$ are fixed to be 256 and 16). The computational domain divisions for all cases with double TITAN XPs are the same as above. Scaling laws with linear fitting lines are plotted in Figure 9. The additional communication time resulted from MPI parallelization is estimated by the difference of the execution time on double GPUs and half of that on single GPU. It is found that the communication time are weakly dependent on the value of $N_R$ though the sizes of communication area are controlled to be identical. With these scaling laws, the parallel efficiency $P$ for a specified MPI domain decomposition with mesh size of ($N_R$, 256, 16) in $R$, $Z$, and $\phi$ can be estimated with

$$P = \frac{3.56N_R + 100.00}{\frac{0.13N_R + 63.92}{256} \times L_{Communication}^{Maximum} + (3.56N_R + 100.00)/N_{GPUs}} \times \frac{1}{N_{GPUs}} \quad (15)$$

where the numerator estimates the execution time on single TITAN Xp, while the denominator estimates that on $N_{GPUs}$ TITAN Xps. The first term in denominator estimates the communication time (the communication time is assumed to be proportional to the maximum length of grids $L_{Communication}^{Maximum}$ in all processors used for communication, for example, $L_{Communication}^{Maximum}$ equals to 256 in all cases of Figure 9). And the second term in denominator estimates the computation time, which is assumed to be inversely proportional to the number of GPUs. With Equation (15), the estimated parallel efficiencies for four cases in Figure 9 are 61%, 69%, 78%, 84% and 88% respectively for $N_R$ equals to 32, 64, 128, 256 and 512, while the actual efficiency values are 55%, 68%, 81%, 84% and 88%. Consequently, the estimation formula is relatively accurate though there is still some error. As for the mesh size of (256, 256, 16) in $R$, $Z$, and $\phi$ used in simulation, the parallel efficiency for 4 GPUs parallelization is estimated to be 72%, and that value drops to a low value of 51% for 8 GPUs. Therefore, the optimal number of GPUs for mesh size of (256, 256, 16) is expected to be 4 on which hopefully about threefold acceleration could be achieved compared with single GPU.

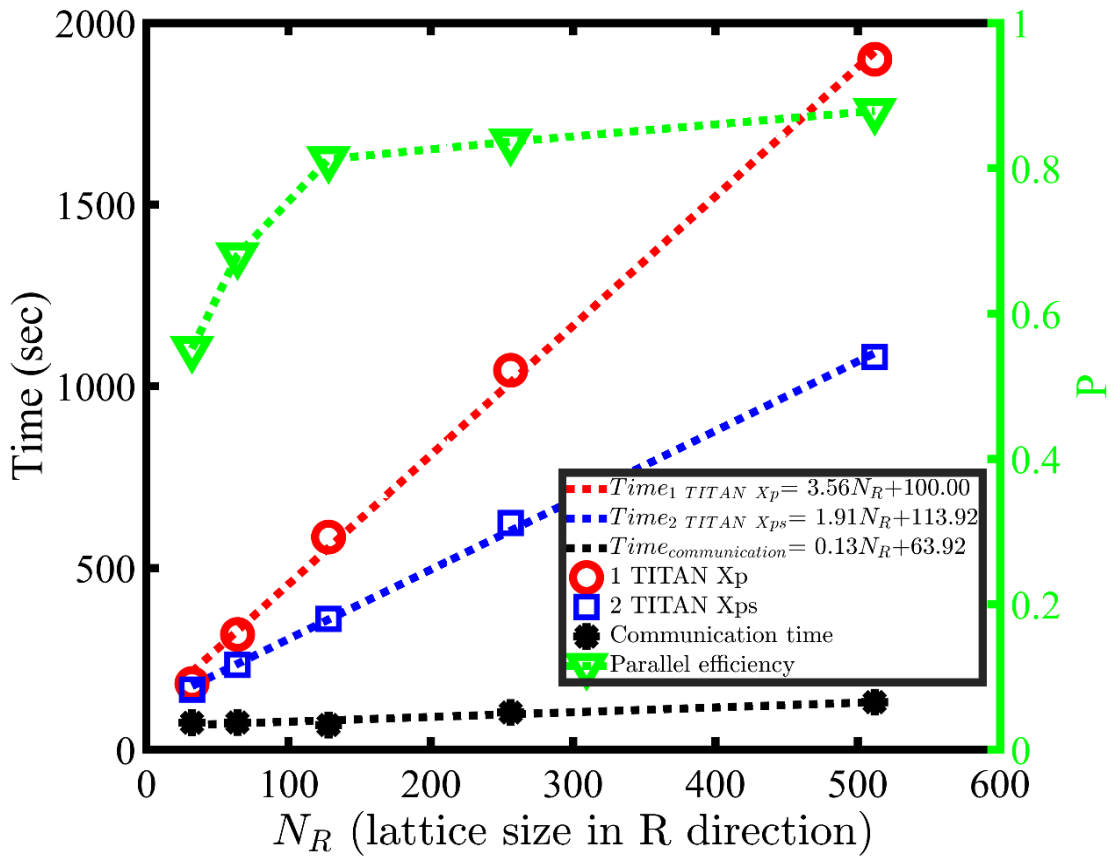

**Figure 9.** Execution times on single or double TITAN Xp GPUs for different mesh size in R direction ($N_R$ equals to 32, 64, 128, 256, and 512, respectively, while $N_Z$ and $N_\phi$ are fixed to be 256 and 16). Red circles indicate the execution times on single TITAN Xp, and blue squares indicate the times on double TITAN Xps. The communication times are plotted with black stars. Linear fitting lines for each case are given. The parallel efficiencies are also plotted with green triangles.

### 4.3 Performance study for the double precision computational capability

Another issue of the acceleration performance of the GPU-OpenACC must be discussed here is the computational capability of the double precision floating points. The theoretical double precision capability of TITAN Xp is up to 0.38 TFLOPS, which is only comparable with one Intel® Xeon® Gold 6148F CPU with 20 cores. However, the execution speed of CLTx on TITAN Xp is about 4 times faster than that on the Gold 6148F CPU with 32 cores parallelization. Although the double precision computational capability of TITAN V is about 6.90 TFLOPS that is more than 10 times of TITAN Xp, the real execution speed on TITAN V only achieves less than two times on TITAN Xp. Therefore, the parallelization performance of the code on different platforms does not entirely depend on the theoretical computational capability. Other factors such as the memory accessing time, partition, bandwidth, I/O situation, compiler optimization options, and the network conditions will also affect the execution speed of the code. To study this unexpected result, a simple demo in double precision as shown in Figure 10 is written to investigate the influence of the computational intensity on acceleration efficiency. One parameter N_CYCLE in this demo is used to control the computational intensity. When N_CYCLE is set to a small value, for example, 1 (which is similar to the situation of CLTx), the demo code is a memory intensive task, which means memory access for data read & write is required every few steps. However, if the N_CYCLE value becomes large, such as 100, the demo code becomes a computationally intensive task. Because after each memory access, the data can be locally used and updated for N_CYCLE times without pausing. By scanning the value of N_CYCLE, the time costs for the iterations and the bandwidth utilization of the device memory (only for GPU cases) are compared for Gold 6148F CPU, TITAN Xp, and TITAN V, respectively. The NVIDIA (R) CUDA command line profiler nvprof is used to obtain the execution time and bandwidth utilization of the kernel in Figure 10. The "dram_read_throughput, dram_write_throughput, and dram_utilization" options in metrics mode of nvprof provide the device memory read & write throughput and the utilization level of the device memory relative to the peak utilization on a scale of low (0) to max (10). The detailed results are presented in Table 2.

As shown in Table 2, when the demo code is a memory intensive task, i.e., N_CYCLE = 1, the speed up ratio of TITAN V and TITAN Xp is close to but slightly greater than that of CLTx, with a value of about 2.103. In the meanwhile, the ratio of sustained bandwidth to theoretical peak bandwidth of TITAN V is about 0.853, and the maximal utilization level of device memory has been reached according to the information provided by nvprof. In this case, compared with TITAN Xp, TITAN V does not show much advantage in spite of its powerful computational capability of double precision due to the bandwidth bound. However, when N_CYCLE is increased to a larger value, for example, N_CYCLE = 100, the code becomes computationally intensive. Correspondingly, the ratio of sustained bandwidth to theoretical peak bandwidth of TITAN V is only about 0.056, and the device memory utilization is in the low level. As a result of the low bandwidth utilization, the speed of TITAN V is more than ten times faster than that of TITAN Xp, which is consistent with the difference of double precision computational capabilities between two GPU cards. Similar analysis for CLTx has been done with nvprof, and the result demonstrates that CLTx is bandwidth limited on TITAN V. Therefore, the bandwidth analysis of Table 2 indicates that TITAN V has a great acceleration potential for CLTx by focusing on the data locality and memory access, which requires more efforts on the deep optimization of the code.

```fortran
PROGRAM DEMO
IMPLICIT NONE
REAL*8, DIMENSION(256,256,16,8) :: A, B, C, D
REAL*8 :: TIME_START, TIME_END, TIME_COST
INTEGER :: JX, JZ, JY, M, I_CYCLE, N_CYCLE
!$acc set device_num(0)

N_CYCLE = 1 ! N_CYCLE CONTROLS THE NUMBER OF CIRCULATIONS
A(:,:,:,:)=1.D-6
B(:,:,:,:)=1.D-6
C(:,:,:,:)=1.D-6
D(:,:,:,:)=1.D-6
!$acc enter data copyin(A,B,C,D)

CALL CPU_TIME(TIME_START)
!$acc kernels default(present)
!$acc loop independent collapse(4)
DO M=1,8
DO JY=2,15
DO JZ=2,255
DO JX=2,255
DO I_CYCLE=1,N_CYCLE
A(JX,JZ,JY,M)=A(JX,JZ,JY,M)+A(JX+1,JZ,JY,M)-A(JX-1,JZ,JY,M)+ &
             B(JX,JZ+1,JY,M)-B(JX,JZ-1,JY,M)+ &
             C(JX,JZ,JY+1,M)-C(JX,JZ,JY-1,M)+ &
             DSIN(D(JX,JZ,JY,M))+DCOS(D(JX,JZ,JY,M))
ENDDO
ENDDO
ENDDO
ENDDO
ENDDO
!$acc end kernels
CALL CPU_TIME(TIME_END)
TIME_COST = TIME_END - TIME_START
PRINT*,'N_CYCLE =', N_CYCLE, 'TIME COST = ', TIME_COST

!$acc exit data delete(A,B,C,D)
END PROGRAM DEMO
```

**Figure 10.** A demo code to investigate the influence of computing intensity on acceleration efficiency. The demo code is memory intensive for small N_CYCLE value while the demo code tends to become compute intensive for large N_CYCLE value.

**Table 2.** Time costs comparisons for the demo code in Figure 10 executed at different platforms by scanning the value of N_CYCLE from 1 to 100. Note that the execution time for the kernel in the GPU case is obtained with nvprof.

| N_CYCLE | TITAN Xp (Theoretical Peak Bandwidth: 547.680GB/s) | | | | CPU |
| --- | --- | --- | --- | --- | --- |
| | Time (sec) | Sustained Bandwidth (GB/s) | Bandwidth Ratio (Sustained/Peak) | DRAM Utilization (Scale from 0 to 10) | Time (sec) |
| 1 | (1.251±0.011)E-03 | 264.549±2.378 | 0.483±0.004 | Mid (6) | 0.107±0.004 |
| 2 | (2.466±0.022)E-03 | 133.297±1.200 | 0.243±0.002 | Low (3) | 0.135±0.005 |
| 5 | (6.098±0.062)E-03 | 53.206±0.008 | (9.715±0.002)E-02 | Low (2) | 0.301±0.005 |
| 10 | (1.225±0.001)E-02 | 26.655±0.005 | (4.867±0.001)E-02 | Low (1) | 0.595±0.010 |
| 100 | (1.196±0.051)E-01 | 3.130±0.282 | (5.716±0.515)E-03 | Low (1) | 6.956±0.047 |
| N_CYCLE | TITAN V (Theoretical Peak Bandwidth: 652.800GB/s) | | | | Speedup Ratio (TITAN V/Xp) |
| | Time (sec) | Sustained Bandwidth (GB/s) | Bandwidth Ratio (Sustained/Peak) | DRAM Utilization (Scale from 0 to 10) | |
| 1 | (5.950±0.009)E-04 | 556.768±0.504 | 0.853±0.001 | Max (10) | 2.103±0.016 |
| 2 | (5.977±0.012)E-04 | 545.165±2.382 | 0.835±0.004 | High (9) | 4.125±0.044 |
| 5 | (6.420±0.018)E-04 | 506.097±0.599 | 0.775±0.001 | High (9) | 9.500±0.104 |
| 10 | (1.025±0.001)E-03 | 317.876±0.481 | 0.487±0.001 | Mid (6) | 11.952±0.014 |
| 100 | (9.394±0.007)E-03 | 36.665±0.053 | (5.617±0.008)E-02 | Low (1) | 12.734±0.547 |

## 5. Results of benchmarking

To verify of the results calculated by the OpenACC version CLTx on the GPU, the simulations of m/n=2/1 resistive tearing mode in Tokamak are carried out both on the CPU-MPI platform (Intel® Xeon® Gold 6148F CPU @ 2.40GHz, 32 cores, ECC memory support) and the GPU-OpenACC platform (TITAN V or TITAN Xp, non-ECC memory support, using one GPU for each test).

The initial equilibrium is calculated by the NOVA code (Cheng and Chance 1987) with the initial safety factor $q$ and pressure $p$ profile as shown in Figure 11. The m/n=2/1 tearing mode is most unstable for this equilibrium. The grid set is (256, 256, 16) in $R$, $Z$, and $\phi$. The resistivity $\eta$ is chosen to be $1\times 10^{-5}$.

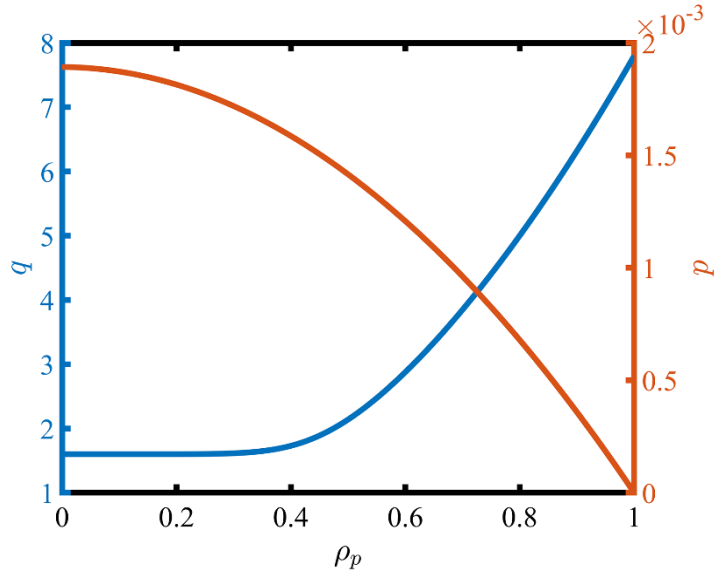

**Figure 11.** Initial equilibrium safety factor $q$ profile and pressure $p$ profile for simulation of m/n=2/1 resistive tearing mode.

Temporal evolutions for the kinetic energy of the system given in Figure 12 are obtained from the CPU-MPI and GPU-OpenACC versions of CLTx. The kinetic energies of the two cases are exactly the same during the whole 400,000 simulation steps, except for some negligible differences on the 14$^{th}$ decimal place of the number, which is almost the machine error limitation of the double precision.

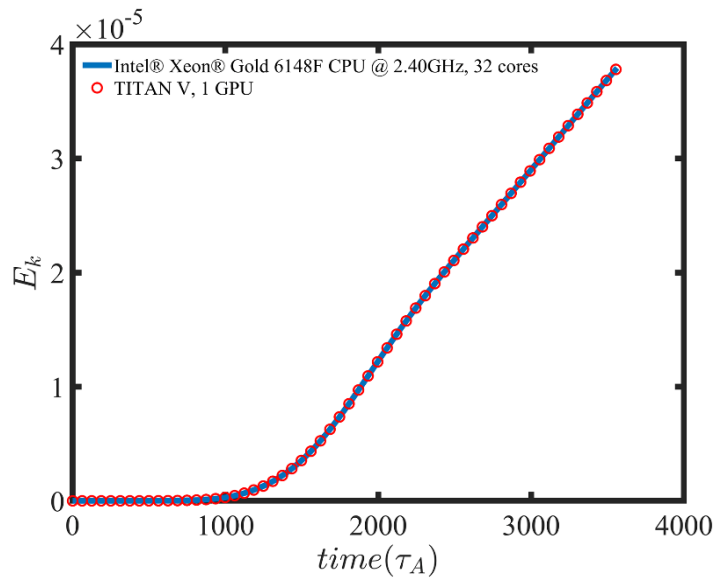

**Figure 12.** Temporal evolutions for the kinetic energy of the system for two cases executed on CPU-MPI platform (blue line) and GPU-OpenACC platform (red circles), respectively.

The mode structures from these two calculations are also identical, as the result of the CPU-MPI case demonstrated in Figure 13(a), and the contour plot for the difference of the mode structures by CPU and GPU is also given in Figure 13(b). The mode structure of the toroidal electric field $E_\phi$ in Figure 13(a) is clearly the m/n=2/1 tearing mode, the result calculated by GPU-OpenACC is omitted due to the indistinguishability. The difference of the results between CPU-MPI and GPU-OpenACC as shown in Figure 13(b) is less than $10^{-14}$, which is consistent with the situation of kinetic energy. As for the result on TITAN Xp, it is just the same as that on TITAN V, thus it will not be repeated here.

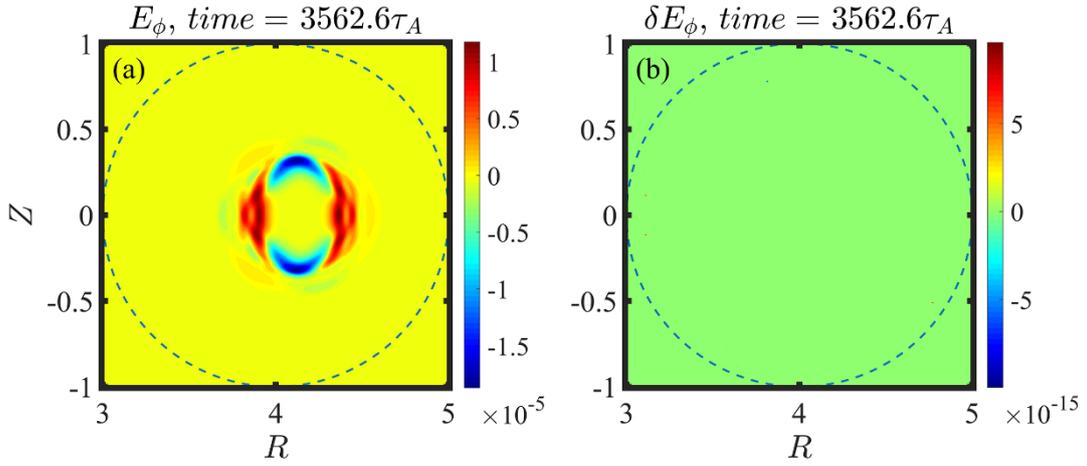

**Figure 13.** (a) The m/n=2/1 tearing mode structure of the toroidal electric field $E_\phi$ and (b) the difference of $E_\phi$ from CPU-MPI platform and GPU-OpenACC platform.

Therefore, the consistency for the m/n=2/1 tearing mode on both platforms confirms the reliability of GPU-OpenACC for the double precision simulation of CLTx. The stability is also verified with dozens of simulations with each over 400,000 steps. The lack of ECC memory on TITAN Xp and TITAN V does not influence the correctness of the CLTx results so far.

## 6. Discussion and Conclusion

The successful application of OpenACC (Kan, Zhang, et al. 2016; Kan, He, Li, et al. 2017; Kan, Lei, et al. 2017; Kan, Liang, et al. 2016; Kan et al. 2018; Kan, He, Ding, et al.

2017) in CLTx leads to great improvement in computational efficiency for studying the MHD instability in Tokamak. The migration of CLTx from the CPU-MPI platform to GPU-OpenACC platform is relatively easy compared with completely rewriting the code into CUDA or OpenCL. Only about 500 lines OpenACC directives have been added into CLTx, which is quite few of lines compared with the total of more than 20,000 lines. Some minor changes have been done on the code, such as merging some small subroutines and adjusting the order of cyclic indices to obtain the best acceleration performance.

Compared with the speed of the same CLTx case executed on CPU with 64 cores MPI-parallelization, about four times of acceleration has been achieved on a single TITAN V and double TITAN Xp GPUs. The combination of MPI and OpenACC makes the multiple GPU acceleration to become feasible by simply using MPI with the data addressed on GPU memory.

It is found that the speedup does not just depend on the theoretical computational capabilities of the CPU and GPU. Only less than double speed of TITAN Xp is achieved on TITAN V for CLTx, while the theoretical computational capability of the latter in double precision is ten times more than that of the former. By increasing the compute intensive of the demo code, the speed of TITAN V is more than ten times faster than that of TITAN Xp. Therefore, greater potential on TITAN V is observed if more optimizations on CLTx, such as increase of data locality, are carried on. However, the speed of the code and the time spent on code modification and optimization is a trade-off. Because the speedup performance on TITAN V is already satisfactory considering its low price, and to keep the integrity and readability of code structure, acceleration work on CLTx is paused at this stage.

Most important of all, the benchmarking for the results calculated by GPU are done by comparing that of traditional CPU platforms. The results from both platforms show exactly identical after 400,000 steps for dozens of runs. The double-precision operations on TITAN Xp and TITAN V can be fully trustable for CLTx.

In addition, the migration to OpenACC for kinetic part of the hybrid kinetic-magnetohydrodynamic version, CLT-K, is still under way, which requires more modifications on the code structure and adjustment on the combination of different OpenACC directives. The experience for this work on CLT-K will be introduced in a future paper if the acceleration performance is noteworthy.

As magnetic confinement fusion is approaching the era of burning plasma, for instance, the under-construction project "International Thermonuclear Experimental Reactor (ITER)" and the under-design project "China Fusion Engineering Test Reactor (CFETR)", simulation studies become much more complicate and large scale which requires that a simulation code has to be efficient. With the GPU acceleration, the simulation period for a typical MHD case reduces from days to hours, which helps us to understand the physics in Tokamak greatly.

Migration CLTx from a CPU machine to GPU with OpenACC can be taken as a reference for both CFD and MHD codes. The GPU-acceleration can be applied in more fields other than fusion research, such as the geo/astrophysics, bioscience, environics.

**Acknowledgments**

The authors would like to acknowledge helpful suggestions given by professor C. Yang, and the Sunway TaihuLight Supercomputer Team at National Supercomputing Center, Wuxi, China.

This work is supported by the Special Project on High-performance Computing under the National Key R&D Program of China No. 2016YFB0200603, the National Natural Science Foundation of China under Grant No. 11775188 and 11835010, Fundamental Research Fund for Chinese Central Universities, Beijing Natural Science Foundation (8184094), IWHR Research & Development Support Program (JZ0145B022018).**References**

Satoshi Matsuoka. 2016. "Porting and optimizing gtc-p on taihulight supercomputer with sunway openacc." Review of. *HPC China*.

Zhang, W, ZW Ma, and S Wang. 2017. "Hall effect on tearing mode instabilities in tokamak." Review of. *Physics of Plasmas* 24 (10):102510.

Zhang, W, S Wang, and ZW Ma. 2017. "Influence of helical external driven current on nonlinear resistive tearing mode evolution and saturation in tokamaks." Review of. *Physics of Plasmas* 24 (6):062510.

Zhu, J, ZW Ma, and S Wang. 2016. "Hybrid simulations of Alfvén modes driven by energetic particles." Review of. *Physics of Plasmas* 23 (12):122506.

Zhu, Jia, Zhiwei Ma, Sheng Wang, and Wei Zhang. 2018. "Nonlinear dynamics of toroidal Alfvén eigenmodes in the presence of tearing modes." Review of. *Nuclear Fusion*.